\documentclass{iau}
\usepackage{multicol}
\usepackage{titlesec}
\usepackage{amsmath}
\usepackage{graphicx}
\usepackage{multirow}
\newcommand{\Ms}{M$_{\odot}$}

\title{Your Title}

\titlespacing*{\subsection}{0pt}{6pt}{3pt}

\begin{document}

\lefttitle{Sengupta et al}
\righttitle{Massive Stars Across Redshifts in the Era of JWST and Large-Scale Surveys}

\jnlPage{1}{7}
\jnlDoiYr{2021}
\doival{10.1017/xxxxx}

\aopheadtitle{Proceedings IAU Symposium}
\editors{A. Wofford,  N. St-Louis, M. Garcia \&  S. Simón-Díaz, eds.}

\title{Dense Circumstellar Medium around Pulsating Massive Stars Powering Interacting Supernovae}

\author{Sutirtha Sengupta$^{1}$, Das Sujit$^{1,2}$, Arkaprabha Sarangi$^{1}$}
\affiliation{$^{1}$Indian Institute of Astrophysics, 
100 Feet Rd, Koramangala, Bengaluru, Karnataka 560034, India.
$^{2}$Pondicherry University, R.V. Nagar, Kalapet, 605014, Puducherry, India.}

\begin{abstract}
We investigate the evolution of red supergiant (RSG) progenitors of core-collapse (CC) supernovae (SNe) with initial masses between $12-20$~\Ms~ focusing on the effects of enhanced mass loss due to pulsation-driven instabilities in their envelopes and subsequent dynamical ejections during advanced stages of nuclear burning. Using time-dependent mass loss from detailed MESA stellar evolution models, including a parameterized prescription for pulsation-driven superwinds and time-averaged mass loss rates attributed to resulting shock-induced ejections, we construct the circumstellar medium (CSM) before the SN explosion. We calculate resulting CSM density profiles and column densities considering the acceleration of the stellar wind. Our models produce episodes of enhanced mass loss $10^{-4}-10^{-2}$~\Ms~$\rm{yr}^{-1}$ in the last centuries-decades before explosion forming dense CSM ($>10^{-15}~\rm{gcm}^{-3}$ at distances $<10^{15}$ cm)  -- consistent with those inferred from multi-wavelength observations of Type II SNe such as SN~2023ixf and SN~2020ywx.

\end{abstract}


\maketitle

\section{~Introduction}

The evolution of massive stars in their late phases remains ill understood - in particular, the effects of mass loss on the final fate of these stars \citep{2014ARA&A..52..487S,2015A&A...575A..60M}. A number of factors have been invoked to explain mass loss from massive stars like stellar winds, rotation, radial pulsations, episodic ejections/eruptions and binarity \citep{2002RvMP...74.1015W,2012ARA&A..50..107L}, which result in the observed diversity of CCSNe, which are broadly classified as Type I (b/c) -- which do not show any hydrogen lines in their spectra -- and Type II (P/L/n/b) which do so, but with varying strengths and time evolution of light curves \citep{1997ARA&A..35..309F}.
There is a growing consensus that a high fraction of Type II SNe are preceded by phases of heavy mass loss \citep{10.1093/mnras/staf888}, with visible signatures in their post-explosion shock interaction with the circumstellar medium (CSM) created by the wind of the SN progenitor \citep{universe11050161,2025arXiv251020913C}. The circumstellar densities infered from flash-ionization spectroscopy, X-ray and radio observations of SNe like SN~1998S \citep{10.1046/j.1365-8711.2002.05086.x} and SN~2023ixf \citep{2023ApJ...956L...5B} point toward mass loss rates of $\sim 10^{-3}-10^{-1}$~\Ms~$\rm{yr}^{-1}$ -- far exceeding those associated with steady winds of red supergiants (RSGs) and at odds with latest mass loss prescriptions inferred from large samples of observed RSGs \citep{2020MNRAS.492.5994B,2023A&A...676A..84Y,antoniadis_2024,2024A&A...681A..17D} as well as empirical mass loss rates derived from much smaller samples of stars \citep{1988A&AS...72..259D,1990A&A...231..134N} that are commonly used in stellar evolution codes \citep{2023ApJ...942...69M}.

\section{~Pulsation-driven mass loss from RSGs}
Many RSGs exhibit large-amplitude radial pulsations with periods of several hundred days \citep{2018ApJ...859...73S, 2019ApJS..241...35R}, driven by the so-called $\kappa$-mechanism \citep{1965ApJ...142..868B, 2008A&A...484...29G} - an interplay between excess radiation pressure and gravitational pull on the outer layers of the RSG envelope. These pulsations can grow in amplitude to produce shocks at the stellar surface that may levitate material above the photosphere, reducing the effective gravity and allowing other processes e.g., radiation pressure on dust and molecules \citep{galaxies13040072} to accelerate the material outward. However, the exact mechanism of mass loss due to the complex interplay of pulsations and convection in RSG envelopes remains ill-understood \citep{2010Ap&SS.328..245G,2022ApJ...929..156G,2025arXiv250812486G,ma2025}.
\subsection{~Pulsation-driven super-winds (PDSW)}

\cite{2010ApJ...717L..62Y} first proposed a pulsation-driven wind mass loss prescription for high mass RSGs with high luminosity-to-mass ratios, which they hypothesized could, in principle, achieve the high rates ($\dot M \sim 10^{-2}$~\Ms~$\rm{yr}^{-1}$) required to explain observed properties of CSM around Type IIn SNe \citep{2018ApJ...858...15M}. \cite{2010ApJ...717L..62Y} showed that the growth rate ($\eta$) of the amplitude of the surface velocities increases with $L/M$ ratio and/or decreasing thermal (Kelvin-Helmholtz) time-scale of the envelope ($\tau_{KH, env}$). Using the open-source  MESA \citep{2023ApJS..265...15J} code, we evolve stellar model sequences with initial masses between $12 - 18$~\Ms~ using the PDSW prescription of \cite{2010ApJ...717L..62Y} beyond core helium burning for $\eta>1, ~ T_{\mathrm{eff}}< 4000~\mathrm{K}$, given by: 
\begin{equation} \label{equation : 2}
\mathrm{\dot M} = \eta^{\alpha}\mathrm{\dot M}_{\mathrm{Dutch}},
\end{equation}
where $\alpha$ is a free parameter and $\mathrm{\dot M}_{\mathrm{Dutch}}$ is the Dutch wind scheme of MESA scaled with a factor of $0.2$, for the empirical mass loss rate estimates to be consistent with latest observations of RSGs \citep{2021ApJ...922...55B,2023ApJ...942...69M}. 

\subsection{~Shock-driven dynamical ejections}
\cite{clayton2018a} further extended the pulsation-driven mass loss into the shock-dominated regime with hydrodynamic MESA models showing launch of ejections upon the surface breakout of strong compression shocks, followed by sufficiently fast expansion that raises material on to an escape trajectory. \cite{clayton2018a} proposed the following mass loss prescription for $\log((\mathrm{L}/\mathrm{L}_\odot)/(\mathrm{M}/\mathrm{M}_\odot))\gtrsim4.1-4.15$:
 \begin{equation} \label{equation : 3}
\log(\dot{M}/\mathrm{M_\odot}~\rm{yr}^{-1}) = 5.93\times \log[(L/\mathrm{L}_\odot)/(M/\mathrm{M}_\odot)]-26.6,
\end{equation}
which we employ to model the possibility of such repeating mass ejections in pulsationally unstable RSG envelopes when the luminosity to mass ratio of the star exceeds the cut-off value of $\log((L/\mathrm{L}_\odot)/(M/\mathrm{M}_\odot))=4.15$.

\begin{figure}[ht]
        \includegraphics[width=0.49\textwidth]{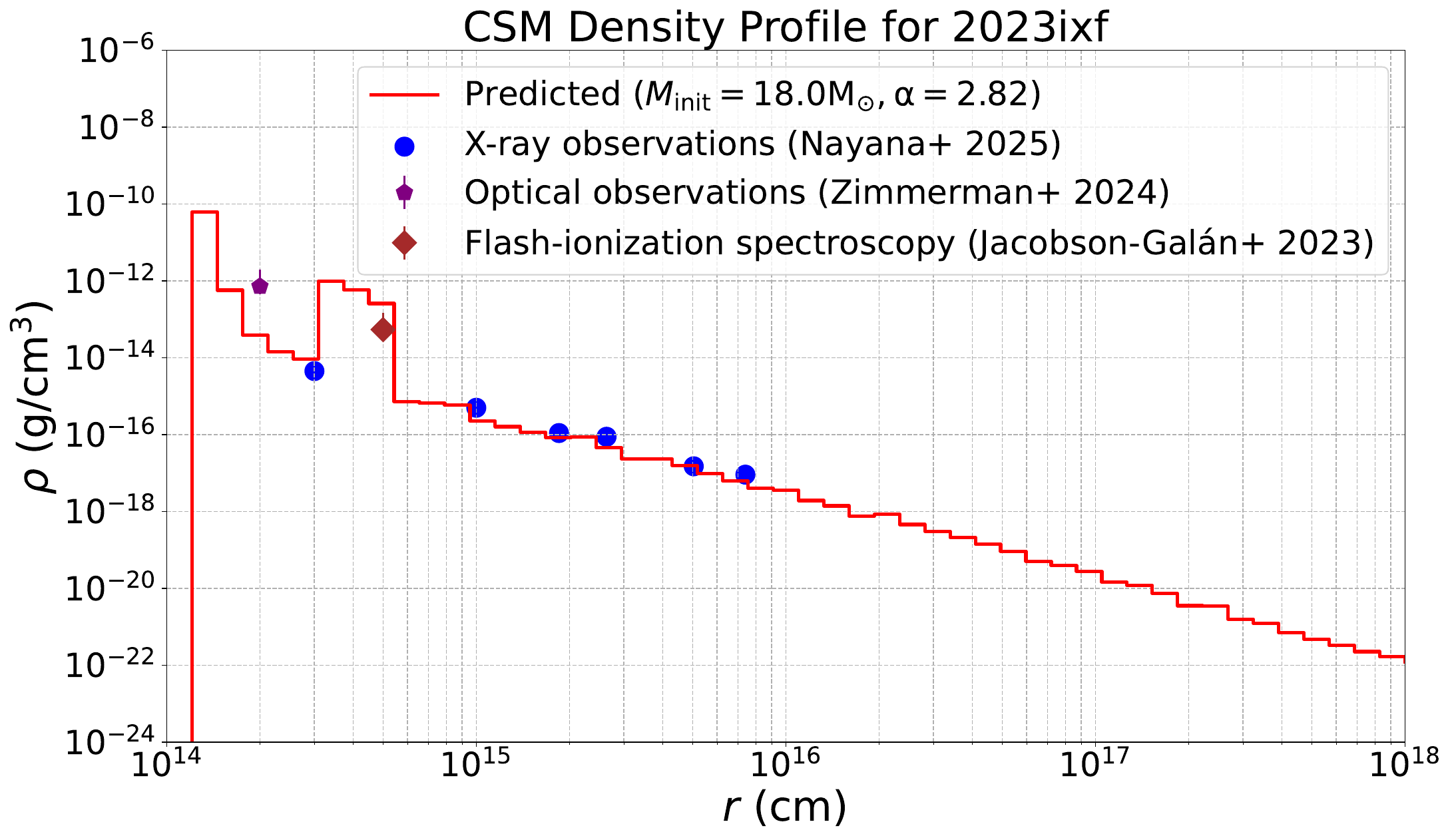}        
        \label{fig:10b}
    \hfill
        \includegraphics[width=0.49\textwidth]{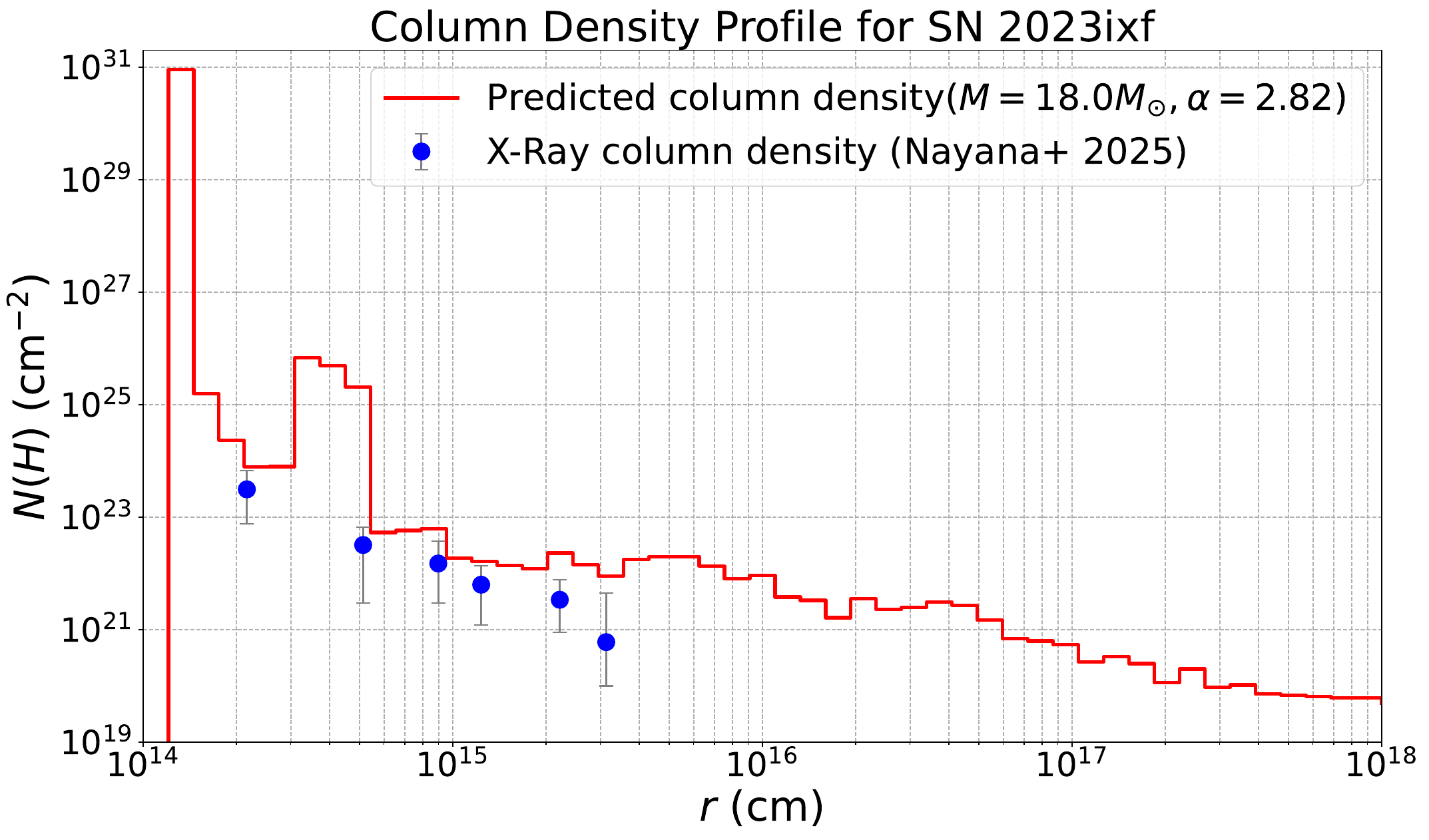}
        \label{fig:10c}

    \caption{(left) CSM density profiles for the model sequence with $M_{init}=18~\rm{M}_\odot,~\alpha=2.82$ (solid red curve), compared with inferred densities from multi-wavelength observations \citep{2023ApJ...954L..42J,2024Natur.627..759Z,2025ApJ...985...51A} and (right) the column density of neutral hydrogen (solid red) compared to X-ray measurements (in solid blue) for SN~2023ixf \citep{2025ApJ...985...51A}.}
    \label{fig:10}
\end{figure}

\begin{figure}[ht]
\centering
        \includegraphics[width=0.49\textwidth]        {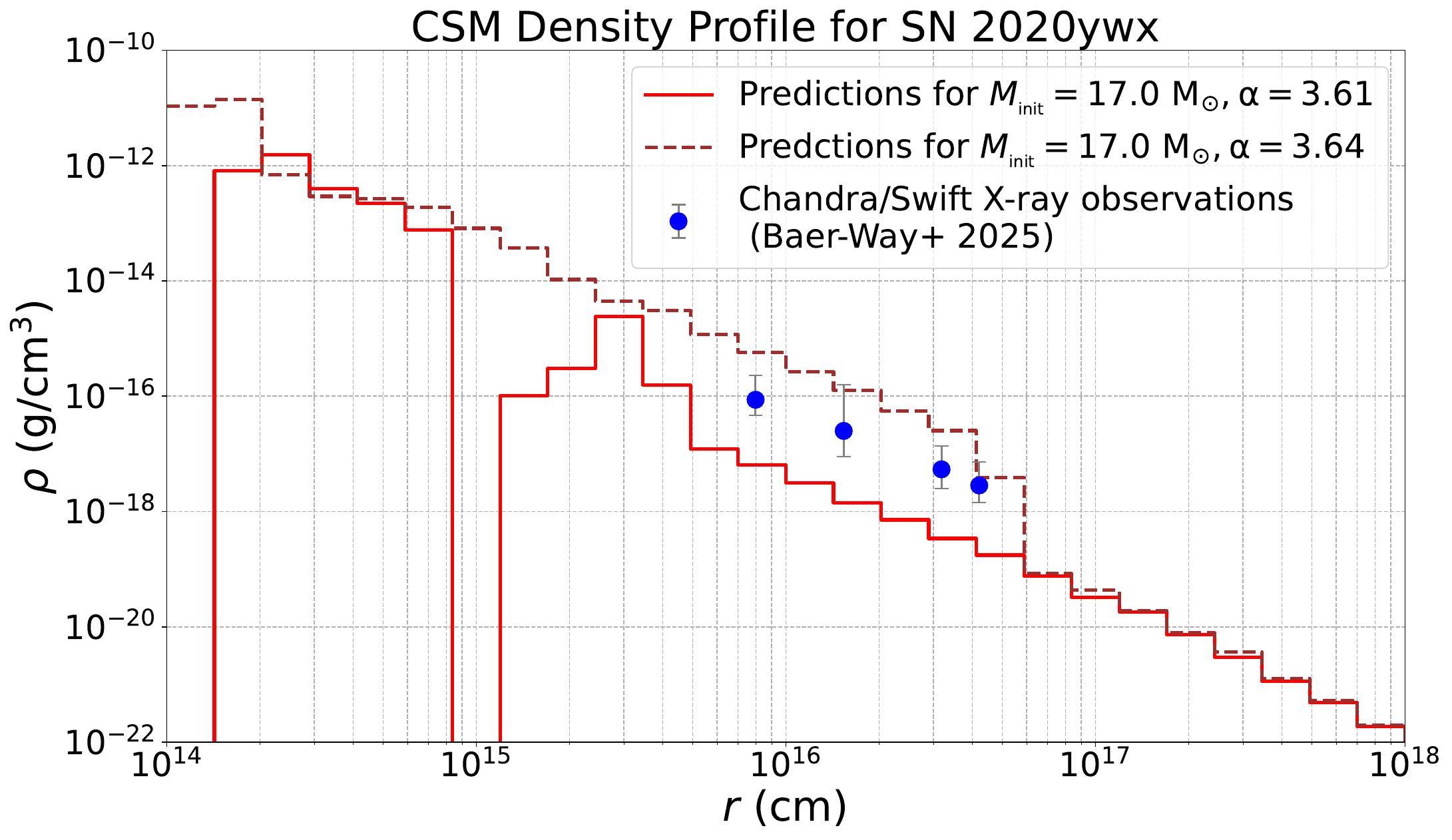}
        \label{fig:9b}
    \hfill
        \includegraphics[width=0.49\textwidth]{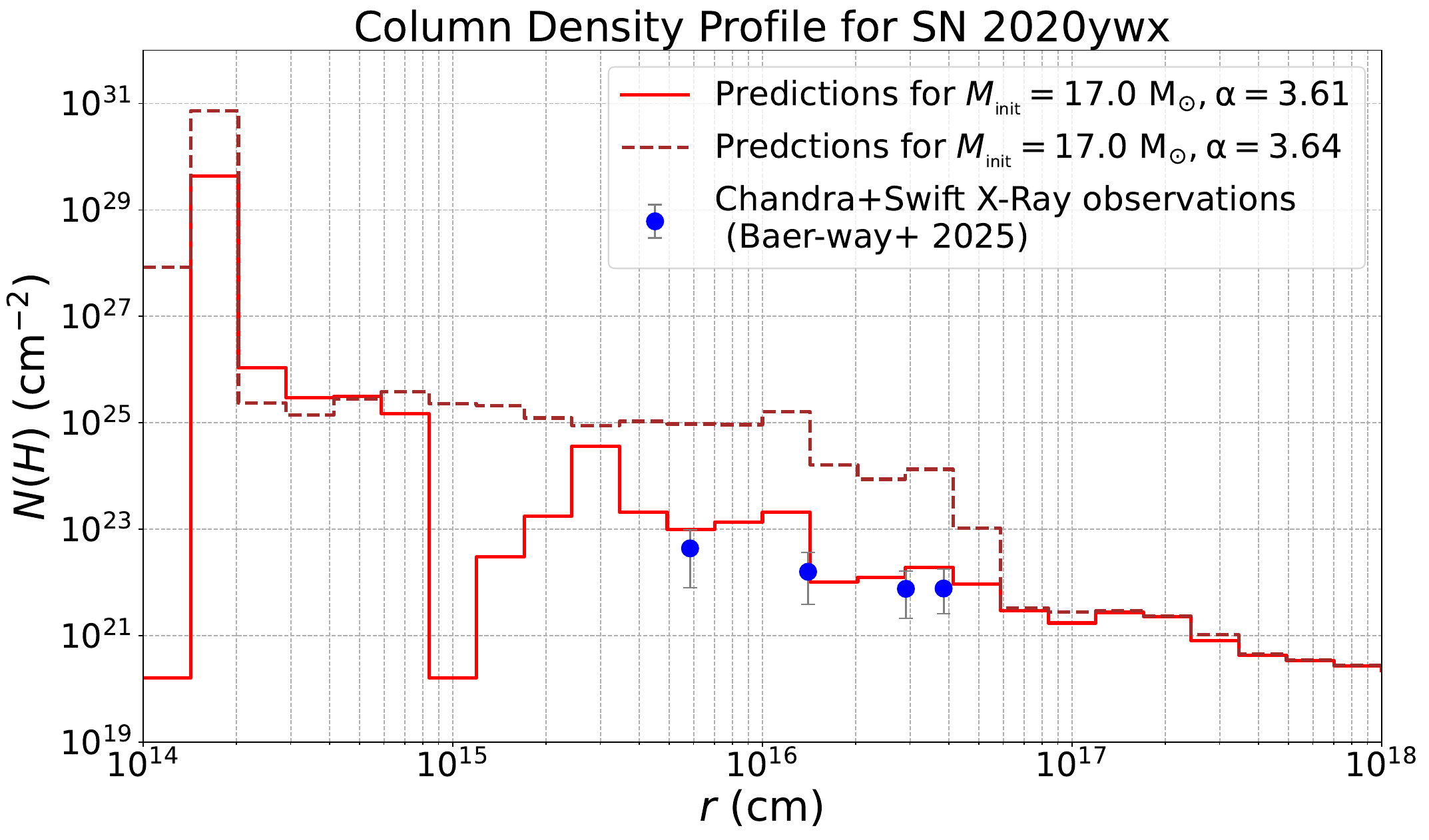}
        \label{fig:9c}
        \caption{(left) CSM density profiles for the two MESA model sequences for $M_{init}=17~\rm{M}_\odot$, with $\alpha=3.61$ sequence given by the solid red curve and the $\alpha=3.64$ sequence given by the dashed brown curve, compared with estimates from measured X-ray luminosities and (right) the column density of neutral hydrogen calculated for the same model sequences, compared to the measured X-ray column densities (blue points) from observations \citep{Baer-Way_2025}.}
    \label{fig:9}
\end{figure}

\section{~Comparison with multiepoch observations of Type II SNe}
We calculate the CSM density and column density of neutral hydrogen along the line of sight assuming spherical symmetric mass loss and an accelerated wind profile from the RSG, given by a simple $\beta$-law \citep{2018MNRAS.476.2840M}, with $\beta=1.2$ -- consistent with observations of single RSGs (\citealt{2010ASPC..425..181B}). Our model results are in good agreement with observations of Type II SNe like 1998S, 2005ip, 2017hcc \citep{2025arXiv250804497S} as well as multi-epoch observations of more recent events like SNe 2023ixf (see Figure~\ref{fig:10}) and SN 2020ywx (see Figure~\ref{fig:9}), although some systematic differences are seen in the column densities which could arise due to explosion asymmetry \citep{2025A&A...694A.319K} or clumping along line of sight effects \citep{Baer-Way_2025}.

\setlength{\bibsep}{0pt} 
\begin{multicols}{2}

\bibliographystyle{iaulike_notitles_all}

\bibliography{Bibliography_sarangi,bibliograhy}
\end{multicols}
\end{document}